\documentclass[twocolumn,showpacs,preprintnumbers,amsmath,amssymb]{revtex4}
\usepackage{psfig}% Include figure files
\usepackage{dcolumn}% Align table columns on decimal point
\usepackage{bm}% bold math
\usepackage{multirow}
\usepackage[T1]{fontenc}
\usepackage{ae,aecompl}
\usepackage{amssymb}
\usepackage{amsmath}
\usepackage{mathrsfs}

\newcommand{\tgg}{\mathscr{G}}

\newcommand{\tgf}{\mathscr{F}}

\newcommand{\cpsi}[2]{\hat{\psi}^\dag_{#1}({#2})}
\newcommand{\dpsi}[2]{\hat{\psi}_{#1}({#2})}
\newcommand{\up}{\uparrow}
\newcommand{\down}{\downarrow}

\newcommand{\bx}{{\bf x}}

\begin{document}

\title{Quasi-particle properties and Cooper pairing in trapped
Fermi gases}

\author{L. Giorgetti$^{1,2}$} 
\altaffiliation{Present address: Laboratoire Kastler Brossel, \'Ecole
Normale Sup\'erieure, 24 Rue Lhomond, F-75231 Paris Cedex 05, France.}
\author{L. Viverit$^{1,2}$} 
\author{G. Gori$^{1}$}
\author{F. Barranco$^3$} 
\author{E. Vigezzi$^{1}$} 
\author{R. A. Broglia$^{1,4}$}
\affiliation{ $^1$Dipartimento di Fisica-Universit\`a di Milano and
INFN Sezione di Milano, Via Celoria 16, I-20133 Milan, Italy\\
$^2$CRS-BEC INFM and Dipartimento di Fisica, Universit\`a di Trento,
I-38050 Povo, Italy\\ $^3$Escuela de Ingenieros Industriales,
Universidad de Sevilla, Camino de los Descubrimientos, Sevilla,
Spain\\ $^4$The Niels Bohr Institute, University of Copenhagen,
Blegdamsvej 17, DK-2100 Copenhagen, Denmark}

%\bigskip
%\bigskip
\begin{abstract}
The possibility for the particles in a Fermi gas to emit and reabsorb
density and spin fluctuations gives rise to an effective mass and to a
lifetime of the quasi-particles, as well as to an effective pairing
interaction which affect in an important way the BCS critical
temperature.  We calculate these effects for a spherically symmetric
trapped Fermi gas of $\sim$ 1000 particles.  The calculation provides
insight on the many-body physics of finite Fermi gases and is closely
related to similar problems recently considered in the case of atomic
nuclei and neutron stars.
\end{abstract}



\maketitle
%\newpage

The experimental advances in cooling and manipulation of atomic Fermi
gases are not only important in their own right, but also because of
the possibility of providing a deeper understanding of the physics of
finite many-body Fermi systems.  A crucial feature of atomic Fermi
gases is the possibility of exploiting Feshbach resonances.  These
resonances occur when, due to the action of an external magnetic
field, a two-body bound state crosses threshold.  The result is a
change in the sign of the two-body scattering length $a$ and allows to
explore Fermi superfluidity in all regimes from the weak-coupling
limit, for $a$<0 and $k_F|a|\ll 1$ (with $k_F$ being the Fermi
wavenumber) to the strong-coupling one, for $a$<0 and $k_F|a|\sim 1$,
as well as Bose-Einstein condensation of two-fermion dimers (for
$a$>0) \cite{theory}. This crossover from fermionic to bosonic
superfluidity has been the object of extensive work in the recent
years and the first experimental data are being collected
\cite{expts}.  A fundamental theoretical problem is the evaluation of
the critical temperature and the $T=0$ equation of state in the whole
crossover region.  In this paper we concentrate on the calculation of
the effective quasi-particle interaction, mass and lifetime in a
trapped Fermi gas as the resonance is approached from the fermionic,
$a<0$, side.  As one moves closer to resonance the strength of the
effective interaction increases and the quasi-particle properties are
strongly renormalized due to the possibility of emitting and
reabsorbing density and spin fluctuations.  This problem is
particularly important when studying fermionic superfluidity, since
both the superfluid critical temperature, $T_c$, and the amplitude of
the $T=0$ pairing gap depend strongly on the actual quasi-particle
properties.  These features are normally neglected in the standard
crossover theories. As we shall show, they have a strong effect on
$T_c$.  In what follows we develop a framework to take into account
the effective quasi-particle properties and which is appropriate to
finite size systems in a non-uniform isotropic trap. The case of
uniform atomic Fermi gases has been dealt width in
Ref. \cite{COMBESCOT}.  In what follows we show that the discrete
level structure has a crucial influence on the problem, and point out
the specific effects on $T_c$ of the various quasi-particle
properties. We relate our approach to other calculations carried out
for different Fermi systems. In fact the role of polarization effects on
the quasi-particle properites has been recently studied in a number of
fields of physics as nuclear physics, neutron stars and $^3$He
(cf. e.g. \cite{nuclei,MAHAN} and references therein).

We consider a Fermi gas of $N$ particles in two internal states
described by the Hamiltonian
\begin{eqnarray}
  \nonumber
  \hat{H}&=&\sum_{\sigma=\up,\down}\int d^3 r\;
  \hat{\psi}^{\dagger}_{\sigma}({\bf r})(H_0 -\mu)\hat{\psi}_{\sigma}({\bf
    r})\\ &+&g\int d^3 r\;
  \hat{\psi}^{\dagger}_{\up}({\bf r})
  \hat{\psi}^{\dagger}_{\down}({\bf r})
  \hat{\psi}_{\down}({\bf r})
  \hat{\psi}_{\up}({\bf r})
  \label{Eq:Hamil}
\end{eqnarray}
where $H_0=-\hbar^2\nabla^2/2m_a+V_{ext}(r)$ with
$V_{ext}(r)=m_a\omega_0r^2/2$, $\omega_0$  being the trap frequency,
and $m_a$ the atomic mass. We are
here interested in the BCS side of the resonance where 
$g<0$. We have furthermore assumed the number of particles and
the external confining potentials $V_{ext}$ to be equal for both
internal states so that we also have $\mu_\up=\mu_\down=\mu$.

All the properties of the system are described in terms of  the
normal and anomalous propagators, which satisfy the Dyson equations 
(cf. e.g. ref. \cite{MAHAN})
\begin{eqnarray}
\nonumber
 \tgg_{\up\up}(\omega_n) &=&  \tgg^0_{\up\up}(\omega_n) +
\tgg^0_{\up\up}(\omega_n) [
  \Sigma_{\up\up}(\omega_n)\tgg_{\up\up}(\omega_n)\\
  &-& W_{\up\down}(\omega_n)\tgf^\dag_{\down\up}(\omega_n)],
  \label{Dyson1}\\
% \end{eqnarray} 
%\begin{eqnarray}
\nonumber
 \tgf^\dag_{\down\up}(\omega_n) &=&\tgg^0_{\down\down}(-\omega_n)
  [W_{\down\up}^\dag(\omega_n)\tgg_{\up\up}(\omega_n)\\
&+&\Sigma_{\down\down}(-\omega_n)\tgf^\dagger_{\down\up}(\omega_n)],
  \label{Dyson2}
\end{eqnarray}
where $\tgg_{\up\up}(\omega_n)$ stands for
$\tgg_{\up\up}(\bx,\bx',\omega_n)$ and $\tgg_{\up\up}(-\omega_n)$ for
$\tgg_{\up\up}(\bx',\bx,-\omega_n)$, and similarly for the other
propagators. The quantity $\hbar\omega_n=(2n+1)\pi/\beta$ is a fermionic
Matsubara frequency with $\beta^{-1}=k_BT$, $T$ being the temperature
of the system and $k_B$ the Boltzmann constant.  $\Sigma_{\sigma\sigma}$
and $W_{\sigma,-\sigma}$ are the normal and anomalous self-energies
respectively, and the products of propagators and self-energies imply
integration over the internal spatial coordinates. Completely
analogous equations to (\ref{Dyson1}) and (\ref{Dyson2}) are satisfied
by $\tgg_{\down\down}$ and $\tgf_{\up\down}$. The first step to be
taken in carrying out the calculation, concerns the choice of a
specific approximation for the self-energies.  In this work we account
for the possibility for particles to emit and reabsorb density and
spin fluctuations and therefore assume $\Sigma_{\up\up}
(\omega_n)=gn_\down(\bx)\delta(\bx-\bx')+\Sigma^{\rm
ph}_{\up\up}(\omega_n)$.  The first term corresponds to the usual
$\omega$-independent Hartree self-energy. Treating the
phononic contribution $\Sigma^{\rm ph}_{\up\up}$ to the self-energy in 
the Random Phase Approximation (RPA) one finds: $\Sigma^{\rm
ph}_{\up\up}(\omega_{n})=-(g^2/4\beta)\sum_{m}\tgg^0(\omega_{n}+\omega_{m})
[\Pi_\rho(\omega_{m})+3\Pi_{\sigma_z}(\omega_{m})-2\Pi_0(\omega_{m})]$.
Here $\Pi_0$ is the density correlation function in the Hartree
approximation, while $\Pi_\rho$ and $\Pi_{\sigma_z}$ are the density
and spin correlation functions respectively in the RPA.
The relation between these functions follows introducing  
the Hartree correlation function for a single species $\chi_0$, 
to find $\Pi_0=2\chi_0$ together with $\Pi_\rho=2\chi_0/[1-g\chi_0]$ and
$\Pi_{\sigma_z}=2\chi_0/[1+g\chi_0]$.  Due to the isotropy of the
system the spin-spin correlation functions in the $x$, $y$ and $z$
direction are equal, a fact which is accounted for by multiplying 
$\Pi_{\sigma_z}$ by a factor of 3. Finally we also used
$\tgg^0_{\up\up}=\tgg^0_{\down\down}=\tgg^0$. The Hartree density
correlation function admits the Lehmann representation

\begin{equation}
  \Pi^0(\bx,\bx',\omega_m)=\int_{-\infty}^{+\infty}\frac{d\omega'}{2\pi}\;
  \frac{\pi^0(\bx,\bx',\omega')}{i\omega_m-\omega'},
\end{equation}
with
\begin{eqnarray}
\nonumber
  &&\pi^0(\bx,\bx',\omega')=
  \sum_{ij}\,\frac{e^{-\beta E_i}-e^{-\beta E_j}}{Z}\, \\
  &\times&
  \langle i|\hat\rho(\bx)|j\rangle\langle j|\hat\rho(\bx')|i\rangle
  \delta(\hbar\omega'-E_i+E_j).
  \label{pii}
\end{eqnarray}
In Eq. (\ref{pii}) $\hat\rho(\bx)=
\cpsi{\up}{\bx}\dpsi{\up}{\bx}+\cpsi{\down}{\bx}\dpsi{\down}{\bx}$,
while $i$ and $j$ are the many-body eigenstates in the Hartree
approximation (particle-hole excitations), $E_i$ and $E_j$ are the
corresponding energies and $\omega_m$ is a bosonic
Matsubara frequency.  $\Pi_\rho$ admits a similar representation with
$i$ and $j$ replaced by the many-body eigenstates in the RPA
approximation (collective modes). The same is true for
$\Pi_{\sigma_z}$ where in addition $\hat\rho$ must be replaced by
$\hat\sigma_z=\cpsi{\up}{\bx}\dpsi{\up}{\bx}
-\cpsi{\down}{\bx}\dpsi{\down}{\bx}$.  The anomalous self-energy, on
the other hand, is given by $W_{\up\down}(\omega_n)=-\beta^{-1}\sum_m
V^{\rm eff}_{\up\down}(\omega_m)
\tgf_{\up\down}(\omega_{n}+\omega_{m})$, where $V^{\rm
eff}_{\up\down}(\omega_m)=g\delta(\bx-\bx')+(g^2/4)[\Pi_\rho(\omega_m)
-3\Pi_{\sigma_z}(\omega_m)]$ is the effective quasi-particle
interaction including the exchange of density and spin modes.

In a general spherically symmetric system it is convenient to express
the Dyson equations in the Hartree-Fock (spherical) basis of
eigenstates $\phi_\nu(\bx)=R_{nl}^{\rm HF}(r)Y_{lm}(\Omega)$.
Here  $\nu$ stands for
$n,l,m$ (radial, angular momentum and magnetic quantum numbers) and $Y_{lm}$
is a spherical harmonic. The corresponding energy measured from
the Fermi level is denoted by $\xi_\nu=\xi_{nl}$ independent of $m$.  We
shall also omit the spin indices.  Keeping the matrix
elements of the effective interaction only between states connected by
the time reversal $V_{\nu\nu'}^{\rm
eff}(\omega_{m})\equiv\langle\nu\bar\nu|V^{\rm
eff}(\omega_{m})|\nu'\bar\nu'\rangle$ and solving the Dyson equations in
the proximity of $T_c$ one obtains
\begin{equation}
\label{W}
W_\nu(\omega_n)=-\sum_{\nu'}\frac{1}{\beta}\sum_{m}
  V_{\nu\nu'}^{\rm eff}(\omega_{m})\tgf_{\nu'}(\omega_{n}+\omega_m),
\end{equation}
with $\tgf_\nu(\omega_n)=W_{\nu}(\omega_{n})G_{\nu}(\omega_{n})
G_{\nu}(-\omega_{n})$ and
$G_{\nu}^{-1}(\omega_{n})=i\hbar\omega_n-\xi_\nu-\Sigma^{ph}_\nu(\omega_n)$.

Introducing the Lehmann representation for the $\Pi$'s yields
the following representation for the effective interaction:
\begin{eqnarray}
\label{specV}
V^{\rm
  eff}_{\nu\nu'}(\omega_m)=g_{\nu\nu'}-\int_{0}^{\infty}\frac{d\omega}{2\pi}
v_{\nu\nu'}(\omega)\frac{2\omega}{\omega_m^2+\omega^{2}}
\end{eqnarray}
where
$g_{\nu\nu'}=\langle\nu\bar\nu|g\delta(\bx-\bx')|\nu'\bar\nu'\rangle$
is the matrix element of the bare interaction and
$v_{\nu\nu'}(\omega)=(g^2/4)\langle\nu\bar\nu|[\pi_\rho(\bx,\bx',\omega)-
3\pi_{\sigma_z}(\bx,\bx',\omega)] |\nu'\bar\nu'\rangle$. Correspondingly,
after summation over the internal bosonic Matsubara frequencies, one
obtains for the self-energy
\begin{eqnarray}
\nonumber
&&  \Sigma^{\rm ph}_\nu(\omega_n)= \sum_{\nu'}\int_0^\infty
  \frac{d\omega}{2\pi} \;\sigma_{\nu\nu'}(\omega) \\
&\times&  {\left(\displaystyle{\frac{1+n_B(\hbar\omega)-n_F(\epsilon_{\nu'})}
  {i\omega_n-\hbar^{-1}\epsilon_{\nu'}-\omega}
  +\frac{n_B(\hbar\omega)+n_F(\epsilon_{\nu'})}
  {i\omega_n-\hbar^{-1}\epsilon_{\nu'}+\omega}}\right)}
    \label{sigma}
\end{eqnarray}
with
$\sigma_{\nu\nu'}(\omega)=(g^2/4)\langle\nu\nu'|[\pi_\rho(\bx,\bx',\omega)+
3\pi_{\sigma_z}(\bx,\bx',\omega)-2\pi^0(\bx,\bx',\omega)]
|\nu'\nu\rangle$, and $n_B(x)=(e^{\beta x}-1)^{-1}$ and
$n_F(x)=(e^{\beta x}+1)^{-1}$ being the Bose and Fermi thermal
distribution functions respectively \cite{NOTE}.
Eqs. (\ref{W}), (\ref{specV}) and (\ref{sigma})
constitute a set of equations of the type found in the theory of
strong-coupling superconductivity (see for instance
\cite{MAHAN,SCALAPINO}) and should be solved
self-consistently together with the number equation
\begin{eqnarray}
N=\sum_{\sigma=\up,\down}\int d{\bf x}\;\lim_{\eta\to 0}\sum_{n}
\tgg_{\sigma\sigma}({\bf x},{\bf x},\omega_n)e^{i\omega_n\eta},
\label{numbereq}
\end{eqnarray}
which fixes the chemical potential. At low and intermediate
densities one expects the single-pole approximation for
the propagators to be accurate. Introducing this
approximation for the spectral functions of $G$ and $\tgf$ (see
for instance \cite{FETTER,SCHUCK}) one obtains
\begin{eqnarray}
\nonumber && \tgf_\nu(\omega_n)\simeq
-\displaystyle{\frac{Z_\nu^2\,W_\nu(\epsilon_\nu)}{2\epsilon_\nu}} \\
&\times& \left[\displaystyle{
\frac{1}{i|\hbar\omega_n|-\epsilon_\nu+i\gamma_\nu}-
\frac{1}{i|\hbar\omega_n|+\epsilon_\nu+i\gamma_\nu}} \right],
\label{tgf}
\end{eqnarray}
where $\epsilon_\nu=\xi_\nu+{\rm
Re}\Sigma^{ph}(\omega)|_{\hbar\omega=\epsilon_\nu}$ is the renormalized
single-particle energy, $\gamma_\nu={\rm
Im}\Sigma^{ph}(\omega)|_{\hbar\omega=\epsilon_\nu-i\eta}$ the level width,
and $Z_\nu=(1-\hbar^{-1}\partial\Sigma^{ph}_\nu(\omega)/\partial
\omega|_{\hbar\omega=\epsilon_\nu})^{-1}\leq1$ the quasi-particle
strength. The single-pole approximation is accurate as long as the
resulting single-particle level width is small compared with the
Hartree level spacing and $Z_\nu$ is not too small. At these densities
one also expects the effect of the self-energies $\Sigma^{ph}$ and $W$
on the number equation to be small compared with the Hartree one. We
thus approximated $\tgg_{\sigma\sigma}$ in Eq. (\ref{numbereq}) with
the Hartree expression $\sum_\nu|\phi_\nu({\bf
x})|^2/(i\hbar\omega_n-\xi_\nu)$.  Calculations which go beyond the
one pole approximation \cite{TERASAKI} and which include vertex
corrections \cite{PACO} have been carried out recently to determine
the pairing gap of superfluid atomic nuclei. These calculations have
shown that one recovers the results of the one-pole approximation for
weak and intermediate coupling.

Several further approximation schemes have been
used in the literature.  Introducing Eq.  (\ref{specV}) and
(\ref{tgf}) into Eq.  (\ref{W}) and neglecting the phonon contribution
to the effective interaction one recovers the approximations of
Ref. \cite{MOREL}. For a Fermi gas this is in general a very poor
approximation since, for instance, in a uniform system the phononic
contribution to the interaction produces a constant reduction factor
$(4e)^{1/3}\simeq 2.2$ in the critical temperature in the
weak-coupling limit \cite{GORKOV}. On the other hand, if one keeps the
phonon induced interaction and neglects the quasi-particle width one
recovers the result of Ref. \cite{SCHUCK}, found in the context of
neutron matter.  As we shall see, it is the quasi-particle width which
has the largest effects on the critical temperature in the
intermediate and strong coupling regimes.  For these reasons we include
both effects, together with the quasi-particle strength and the
renormalization of the single particle energies, to get
%\begin{widetext}
\begin{eqnarray}
\nonumber
\Delta_\nu &=&\sum_{\nu'}
\left[\bar g_{\nu\nu'}-2\hbar\int_0^{\infty}\frac{d\omega}{2\pi}\frac{Z_\nu
v_{\nu\nu'}(\omega)Z_{\nu'}}{|\epsilon_{\nu}|+|\epsilon_{\nu'}|+\hbar\omega}
\right] 
\displaystyle{\frac{\Delta_{\nu'}}{2\epsilon_{\nu'}}}\\
&\times& h(\epsilon_{\nu'},\gamma_{\nu'},T)
 \label{gapfinal}
\end{eqnarray}
%\end{widetext}
with $\bar g_{\nu\nu'}=Z_\nu g_{\nu\nu'} Z_{\nu'}$and
$\Delta_{\nu}=Z_\nu W_\nu(\epsilon_\nu)$. The dimensionless function
$h(\epsilon_{\nu'},\gamma_{\nu'},T)$ is given by \cite{MOREL}:
\begin{eqnarray}
h(\epsilon_{\nu'},\gamma_{\nu'},T)=2k_BT\sum_{n=0}^{\infty}\left[
\frac{2\epsilon_{\nu'}}{(\hbar\omega_n+\gamma_{\nu'})^2
+\epsilon_{\nu'}^2}\right],
\end{eqnarray}
\begin{figure}[htt]
\vspace{.65cm}
   \centerline{\psfig{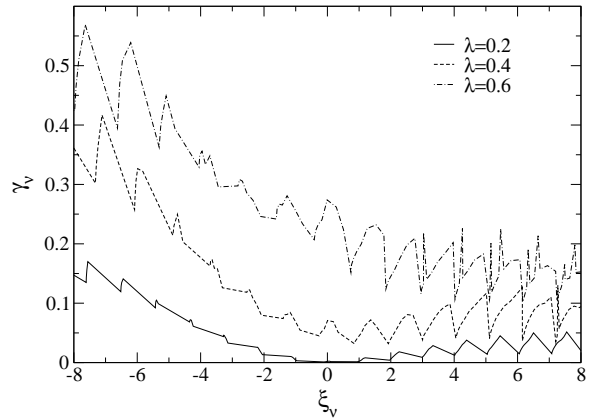}}
    \caption{ $\gamma_\nu$ as a function of the Hartree quasi-particle energy
      $\xi_\nu$ for $\lambda=0.2$ (solid line), $\lambda=0.4$
      (dashed line) and $\lambda=0.6$ (dot-dashed line).  The
      quantities are measured in units of $\hbar\omega_0$.}
\label{fig:imtrap}
\end{figure}
\begin{figure}[hhh]
\vspace{.55cm}
    \centerline{\psfig{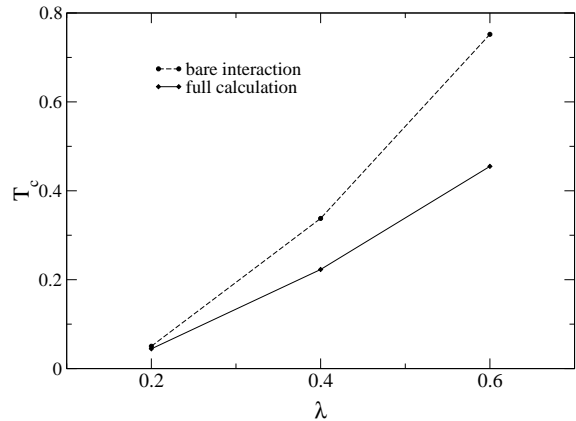}}
    \caption
      {$T_c^{\rm bare}$ and $T_c$ (in units of $\hbar\omega_0$) as a function
       of the interaction strength $\lambda$. The values of $T_c^{\rm
       bare}$ are given in the caption of Table I, while those of
       $T_c$ include all contributions in Eq. (\ref{gapfinal}).}
  \label{fig:Tc}
\end{figure}
and reduces to the usual hyperbolic tangent of BCS theory when
$\gamma_{\nu'}=0$.  The effective interaction is here treated in the
Bloch-Horowitz approximation.  This means that we have neglected the
effect of the level width on the induced interaction and taken the
absolute value of the external quasi-particle energy in the
denominator. A detailed comparison between the results
of this approximation and of the full self-consistent Gorkov theory
was carried out in Ref. \cite{TERASAKI}. There it was shown that the
two approaches lead to very similar results in the weak and
intermediate coupling regimes. Consequently we kept this level of
approximation in our calculations. Eq. (\ref{gapfinal}) is an
eigenvalue equation for $\Delta_{\nu}$ and $T_c$ is the highest
temperature at which the eigenvalue 1 appears. Notice that the
equation contains a sum over $\nu'$ which needs special care.  The
contribution of the bare interaction, in fact, leads to an
ultra-violet divergence, since it arises from a contact approximation
to the true interatomic potential. Several approaches have been
developed to eliminate this divergence, approaches which involve a
renormalization of the coupling constant $g$ or the introduction of a
pseudopotential \cite{RENORM}.  For a trapped gas, for which only the
bare interaction is considered, the results of these calculations may
be reproduced by introducing a cut-off in the sum at
$\epsilon_{\nu'}=\epsilon_F$ \cite{BRUUN}. Because the main scope of
the present work is on the phonon induced effects, we have treated the
bare interaction at this level of approximation.  
Due to the explicit dependence on $\epsilon_{\nu'}$, the phonon
induced interaction has a natural cut-off. Numerical
calculations have shown that the terms with
$|\epsilon_{\nu'}|\gtrsim\epsilon_F$ give negligible
contributions.

Applying the formalism to a spherically symmetric infinite square well
we found the results of Gorkov and Melik-Barkhudarov \cite{GORKOV} in
the weak-coupling limit. In the intermediate coupling regime we did not
find the strong reduction predicted in Ref. {\cite{COMBESCOT}; this
may be due to the different set of approximations used.
For a harmonically confined gas the finite level spacing
strongly affects the properties of the system, and we find the
results reported in Table I. We have performed the calculations
for a cloud of $\sim1000$ particles and for three different
coupling strengths $\lambda$. Here  $\lambda$ is defined in
analogy with a uniform system as $|g| m_a k_F/2\pi^2\hbar^2$ with
$k_F=(3\pi^3)^{1/3} [n(0)]^{1/3}$, and $n(0)$ being the central
density of the cloud.

\begin{table}
\begin{tabular}{|c|ccc|}
\hline
& \;\;\; $\lambda$=0.2\;\;\; &\;\;\; $\lambda$=0.4\;\; \;&
\;\;\; $\lambda$=0.6\;\;\;\\
\hline \hline
$V^{\rm eff}$ & 0.92  &  0.86 & 0.90\\
\hline
$g$+ Re$\Sigma^{\rm ph}$ & 1.01 & 1.03 & 1.03\\
\hline
$g$+ Z & 0.98  & 0.90 & 0.93\\
\hline
$g$+ $\gamma$ & 0.97  & 0.85 & 0.76\\
\hline
$V^{\rm eff}$+ Re$\Sigma^{ph}$ + $\gamma$ + Z & 0.88  & 0.66 & 0.61 \\
\hline
\end{tabular}
\caption{The table gives the ratio $T_c/T_c^{\rm bare}$ associated with
the indicated
contribution. $T_c^{\rm bare}$ is
calculated including only $g$ and we find the following results: for
$\lambda=0.2$,  $k_BT_c^{\rm bare}=5.03\times10^{-2}\hbar\omega_0$, for
$\lambda=0.4$, $k_BT_c^{\rm bare}=0.34\hbar\omega_0$ and for $\lambda=0.6$,
$k_BT_c^{\rm bare}=0.75\hbar\omega_0$.}
\end{table}

In the weak-coupling regime the effects of the induced interaction and
self-energy are strongly suppressed by the discrete shell structure,
and the reduction predicted by Gorkov \cite{GORKOV} for a uniform
system is almost absent. This is because the largest contributions to
the sum in Eq. (\ref{gapfinal}) comes from the condition
$\epsilon_\nu=\epsilon_{\nu'}=\omega=0$, which corresponds to
particles at the Fermi surface and to the exchange of a phonon with
zero energy. The discrete level structure causes the values of the
spectral functions relative to the $\Pi$ functions (see
Eq. (\ref{pii})) to be negligible near $\omega=0$, and consequently
the effect of the phonon-induced interaction to be small.

The real part ${\rm Re}\Sigma^{\rm ph}$ causes an increase in the
critical temperature since it leads to an effective increase of the
density of levels at the Fermi energy, and thus to a stronger
effective coupling strength. On the other hand, both $Z_\nu$ and
$\gamma_\nu$ depress the value of $T_c$.  The former causes a weaker
effective interaction between quasi-particles, as can be deduced from
Eq. (\ref{gapfinal}) recalling that $Z_\nu\leq 1$ \cite{notere}.
Because $\gamma_\nu$ measures the energy range over which the
quasi-particle state is spread due to the coupling to vibrations
(lifetime), its presence effectively inhibits pairing between
quasi-particles \cite{MOREL}.  In Fig.~\ref{fig:imtrap} we show the
value of $\gamma_\nu$ for three different coupling strengths.  Its
effect is important in the intermediate coupling regime, as can be
seen in Table I for the cases $\lambda=0.4$ and $\lambda=0.6$.  In
Fig. \ref{fig:Tc} are shown the values of $T_c^{\rm bare}$, obtained
when only the (bare) direct interaction $g$ is included, and of $T_c$
when all the terms in Eq. (\ref{gapfinal}) are considered.

We acknowledge useful discussions with P.F. Bortignon and the
technical support of F. Marini.


\begin{thebibliography}{100}
\bibitem{NOTE} The evaluation of $v_{\nu,\nu'}$ and
$\sigma_{\nu,\nu'}$ in Eqs. (\ref{specV}) and (\ref{sigma}) involves
expansion in multipolarities of these quantities. We found that for
$\sim 1000$ particles multipolarities up to $L_{\rm max}\simeq 25$
contribute. The scale of $L_{\rm max}$ can be estimated imposing the
associated wavelength to be smaller than the interparticle spacing
$[n(0)]^{-1/3}$.
\bibitem{notere} Because of the discrete level
structure, $\Sigma^{ph}_\nu(\omega)$ is not a continuous function of
$\omega$. One should therefore interpolate it with a continuous one
to calculate $Z_\nu$.
\end{thebibliography}
\end{document}